\documentclass[aps,preprint,superscriptaddress,showpacs,floatfix]{revtex4}%
\usepackage{amsfonts}
\usepackage{amsmath}
\usepackage{amssymb}
\usepackage{graphicx}
\usepackage{graphicx}%
\begin{document}
\title{Application of M\"{o}ssbauer spectroscopy to study vibrations of a
granular medium excited by ultrasound}
\author{R. N. Shakhmuratov}
\affiliation{Zavoisky Physical-Technical Institute, FRC Kazan Scientific Center of RAS,
Kazan 420029 Russia}
\affiliation{Kazan Federal University, 18 Kremlyovskaya Street, Kazan 420008 Russia}
\author{F. G. Vagizov}
\affiliation{Kazan Federal University, 18 Kremlyovskaya Street, Kazan 420008 Russia}
\date{{ \today}}

\begin{abstract}
It is shown that analysis of M\"{o}ssbaur spectra of a granulated medium,
immersed into the epoxy resin without hardener, allows to study mechanical
vibrations of granules. In our experiments, small particles of the potassium
ferrocyanide with a 1.25 micron mean size played a role of granules. This compound
was enriched by $^{57}$Fe isotope. Particle vibrations in the vibrated resin with the
frequency 12.72 MHz were induced by piezo polymer film. At rest, M\"{o}ssbauer
spectrum of $^{57}$Fe in the potassium ferrocyanide consists of a single line.
Ultrasonic vibration of nuclei splits the line into a comb structure with a period
equal to the vibration frequency. The spectrum analysis allows to estimate the
vibration amplitude of particles and decay of the ultrasound in this medium. The
proposed method is unique since it allows to measure subangstrom displacements of
particles vibrating with several MHz frequency.
\end{abstract}
\maketitle

Properties of microscopic colloidal particles suspended in viscous liquids are of
interest as an example of an overdamped system where viscous forces dominate
over inertial forces, see, for example, Ref. \cite{Berner}. Colloids
play prominent role in biophysics as sensitive probes for studying molecular forces.
Moreover, acoustic attenuation in unconsolidated granular material is of interest.
For example, studying sound propagation in marine sediments is important not only
from the viewpoint of applications, but has a fundamental interest since it gives
information about energy dissipation in a granular medium, see, for example,
Ref. \cite{Buck}. Many textbooks are devoted to the description of dynamics of granular
and colloidal particles, their kinetics and phase transition in these systems,
see, for example, Refs. \cite{Jack,Dhont}.

Experimental tracking of the granular medium dynamics in two-dimensional geometry
was realized optically \cite{Clement}. In three-dimensional case the measurements
were performed by nuclear magnetic resonance imaging \cite{Altob,Nagel}, x-ray
technique \cite{Baxter}, and by tracking a thin layer of radioactive particles
placed between two layers of the test material \cite{Harwood}. Accuracy of these
measurements is not high. Moreover, the listed methods allow to study only low
frequency vibrations of particles. In this paper we report the results
of our study of colloidal particles vibration on ultrasound frequency by
gamma-radiation with 14.4 keV energy. The wavelength of this radiation field
is slightly less that one Angstrom (86 pm). The proposed method allows to
study vibration of colloidal particles with extremely small amplitudes.
Information about attenuation of sound propagating in colloidal medium
along the direction of gamma-radiation contains in the distribution of the
vibration amplitudes along this direction.

This method of the vibration amplitude measurement is based on particular
properties of gamma-radiation interaction with harmonically vibrated nuclei.
We use solid state particles (crystals) containing nuclear isotope $^{57}$Fe,
which has a single absorption line of gamma-radiation at rest, i.e., when
particles do not vibrate. When particles vibrate harmonically with the frequency
$\Omega$, then this line is split into a comb structure with the period $\Omega$
\cite{Ru,Cr,Mi,M77,Sh,SV}.

The nature of this phenomenon is explained as follows. Let a nucleus harmonically
vibrates along the direction of the gamma-radiation propagation and a distance between
the nucleus and the radiation source changes according to the law
$R(t)=R_{0}+ r_{0}\sin(\Omega t+\psi)$, where $R_{0}$ is that part of the distance,
which is fixed, $r_{0}$ is the amplitude, $\Omega$ is the frequency, and $\psi$ is the
phase of the mechanical vibrations. Then, in the coordinate system of the moving nucleus
the radiation field is transformed as
$E_{S}[t, R(t)]=E_{0}(t)\exp {[-i\omega_{S}t +ikR_{0}+i\varphi(t)]}$,
where $E_{0}(t)$, $\omega_{S}$, and $k$ are the amplitude,
frequency and the wave number of the radiation field, respectively,
$\varphi(t)=m\sin(\Omega t+\psi)$ is the periodically oscillating phase,
$m=2\pi r_{0}/\lambda$ is the phase modulation index, and $\lambda$ is the  wavelength
of the radiation field. According to the Jacobi-Anger expansion the radiation field,
$E_{S}[t, R(t)]$, can be expressed as polychromatic field consisting of a set of spectral
lines $\omega_{S} - n\Omega$ ($n=0$, $\pm1$, $\pm2$,...), i.e.,
\begin{equation}
E_{S}[t,R(t)]= E_{C}(t)e^{-i\omega_{S}t+ikR_{0}}
\sum_{n=-\infty}^{+\infty}J_{n}(m)e^{in(\Omega t+\psi)}, \label{Eq1}%
\end{equation}
where $E_{C}(t)$ is the time dependent amplitude of the field, which is the same for
all spectral components, and $J_{n}(m)$ is the Bessel function of the $n$-th order.
The Fourier transform of this field is
\begin{equation}
E_{S}(\omega)=E_{0}e^{ikR_{0}}\sum_{n=-\infty}^{+\infty}J_{n}(m)
e^{i n \psi}E_{0}(\omega_{S}-n\Omega-\omega), \label{Eq2}%
\end{equation}
where $E_{0}(\omega_{S}-\omega)$ is the spectrum of the source field.

Changing the frequency of the radiation source by the Doppler effect, one
can observe the comb structure in the transmission spectrum of the
absorber, harmonically vibrated as a whole. Absorption lines appear each
time when particular frequency $\omega_{S} - n\Omega$ of the comb is
in resonance with a single line absorber with frequency $\omega_{A}$.

In M\"{o}ssbauer spectroscopy the dependence of the number of photon counts
on Doppler shift $\Delta$ gives the spectrum, which for simplicity can be
described as
\begin{equation}
N_{out}(\Delta)=\sum_{n=-\infty}^{+\infty}J_{n}^{2}(m)L_{n}(\Delta), \label{Eq3}%
\end{equation}
where $\Delta=\omega_{S}-n\Omega -\omega_{A}$ is the difference of the
resonant frequency of the single line absorber and the frequency of the spectral
component $\omega_{S}-n\Omega$ of the source in the vibrated reference frame,
$L_{n}(\Delta)$ is the transmission function of the single line absorber.

According to Eq. (\ref{Eq3}), the depth of the $n$-th component of the absorption
line in the transmission spectrum is proportional to the spectral density of the
$n$-th component of the frequency comb since only this component interacts with
the absorber, and the other components propagate through without
resonant interaction. Therefore, the intensity of the transmitted radiation
for the $n$-th resonance is proportional to $1-J_{n}^{2}(m)[1-L_{n}(\Delta)]$.
For the optically thick absorber we have $L_{n}(0) \rightarrow 0$, and the value
of the dip in the transmission spectrum is proportional to $1-J_{n}^{2}(m)$.

The Bessel function $J_{n}(m)$ oscillates with increase of $m$. For example,
for $m=2.4$ the central absorption line must disappear since $J_{0}(2.4)=0$.
Multidirectional evolution of the depths of the absorption lines around the
modulation index value $m=2.4$, i.e., decreasing of the central line with $n=0$
and growing of the first side satellites with $n=\pm 1$, was observed in
Refs. \cite{M77,Sh,SV}. Also, zero spectral density of the central component
of the radiation field ($n=0$) at $m=2.4$ leads to the acoustically induced
transparency of the resonant absorber, which was proposed and experimentally
observed in Ref. \cite{R}.

However, multidirectional dependence (decreasing/increasing) of the spectral
components of the vibrated absorber with increase of the modulation index $m$
was experimentally observed only in bulk solids, for example, in
stainless-steel (SS) foil, which vibrates almost uniformly. Such a medium we name hard.
Vibrations of SS foil were studied in Refs. \cite{Sh,SV}. There, it was shown
that SS foil vibrates as a whole (piston like). Moreover, in Ref. \cite{SV}, the
led mask with a small round hole 0.6 mm in diameter was used. This mask allowed
to track vibrations of small parts of the foil individually demonstrating that
they vibrate almost uniformly.

Absorption spectra of powder samples dispersed in Perspex cement \cite{Cr}
or pressed into a tablet \cite{Sh} look very different. We name such samples
as elastic. Central component of their spectra has always the largest depth,
while the depths of other components monotonously decrease with increasing
number $n$. The spectra of this shape are observed for all values of the
modulation index $m$. Moreover, the number of the observed components increases
with increasing $m$.

These features of the experimentally observed spectra of the elastic media
contradict Eq. (\ref{Eq3}). To explain experimental results the Abragam's model
\cite{A} was applied in Refs. \cite{Cr,Mi,K,C} assuming that nuclei vibrate
independently of each other with random phases $\psi$. The phase randomness
allows to introduce the Rayleigh distribution of the vibration amplitudes of
nuclei. Averaging $\langle J^{2}_{n}(m)\rangle_{R}$ with this distribution changes
the dependence of the $n$-th absorption line in the spectrum, which is described
then by the function
$\langle J^{2}_{n}(m)\rangle_{R}=e^{-\left\langle m^{2}\right\rangle }I_{n}\left(\left\langle m^{2}\right\rangle \right)$,
where $I_{n}\left(\left\langle m^{2}\right\rangle\right)$ is the modified Bessel
function of the $n$-th order and $r_{G}$ is the standard deviation of the random
vibration amplitude from zero value. Qualitatively, this dependence agrees with
experimental spectra of the elastic media experiencing harmonic vibrations.
However, quantitative fitting by this function is poor \cite{Sh}. Moreover,
according to the theory, developed in Ref. \cite{PM}, random phase of nuclear
vibrations must lead to the appreciable decay of the harmonics in time domain
spectra, which was not observed in the experiment \cite{PM}. Random phase
fluctuation of the vibrating nuclei should give also extra broadening of
spectral components of the comb increasing with the number $n$. This follows
from the expression for the $n$-th component of the field,
$e^{i n \psi}E_{0}(\omega_{S}-n\Omega-\omega)$, in Eq. (\ref{Eq2}), where random
phase $n\psi$ is present. Such a broadening is not observed as well.

In this paper we present the results of our new experiments with sodium ferrocyanide
K$_{4}$Fe(CN)$_{6}\cdot$3H$_{2}$O. In the previous report \cite{Sh} we worked with
the pressed powder of this salt. In the experiments, which we report here, powder
crystals of  sodium ferrocyanide were grind up and stirred in epoxy resin without
hardener. The granulated crystals obeyed to lognormal distribution with 1.3 $\mu$m
median size and standard deviation $\sigma=0.18$. The experimental spectra are shown
in Fig 1. They are described well by the averaging of the vibration amplitude distribution
proposed in Refs. \cite{Sh,SV}. This distribution is
\begin{equation}
\langle J^{2}_{n}(m)\rangle_{G}=\frac{\sqrt{\frac{2}{\pi}}%
\int_{0}^{\infty}\exp\left[  -\frac{1}{2}\left(  x-\frac{1}{\varepsilon}\right)
^{2}\right]  J_{n}^{2}(\varepsilon m_{c}x)dx}{1+\operatorname{erf}\left(  \frac
{1}{\sqrt{2}\varepsilon}\right)  } , \label{Eq4}%
\end{equation}
where $m_{c}=2\pi r_{c}/\lambda$, $r_{c}$ is the mean value of the vibration
amplitude and $\varepsilon r_{c}$ is the standard deviation, i.e., $\varepsilon$
describes relative scattering of the amplitudes with respect to its mean value.
In the model \cite{Sh,SV}, it is assumed that nuclei vibrate coherently but with
different amplitudes, which satisfy the Gaussian distribution,
$G_{\textrm{norm}}(r,r_{c})$, centered at the value $r_{c}$ with the standard
deviation $\varepsilon r_{c}$. Taking into account the normalization and
redefying variables and parameters we obtain Eq. (\ref{Eq4}).

Spectrum analysis shows that the scattering of the amplitudes has almost the same
value $\varepsilon= 0.55$ for different amplitudes of the radio frequency (RF)
field. Dependence of the mean value of the modulation index on the voltage of the
RF generator is shown in Fig. 2(a). Modulation index grows almost linearly with
the increase of the RF field amplitude. Distribution function of the vibration
amplitudes is shown in Fig. 2(b) for the RF voltage 8 V and $\varepsilon= 0.55$.
\begin{figure}[ptb]
\resizebox{0.4\textwidth}{!}{\includegraphics{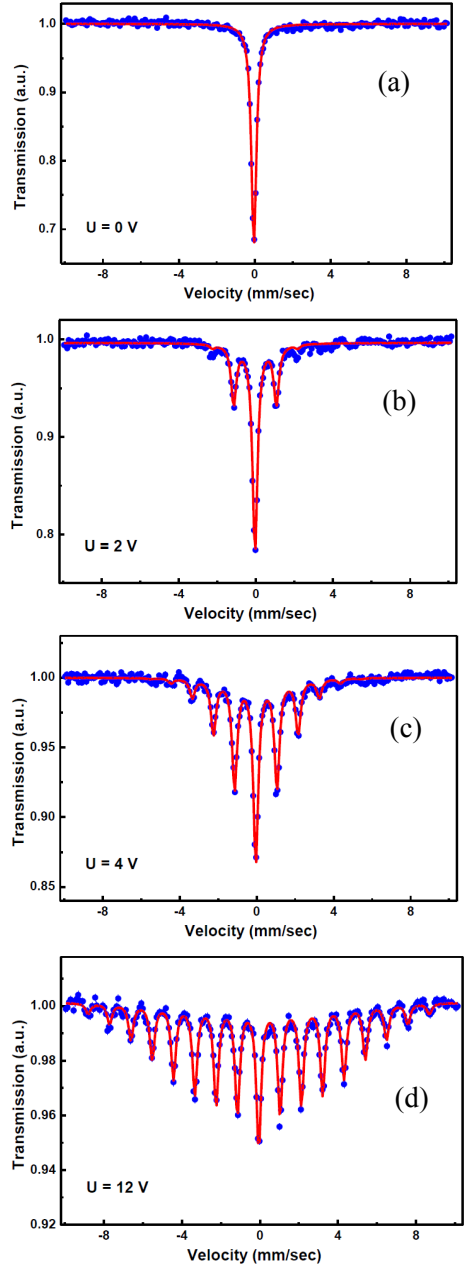}}\caption{Absorption spectra
of the sodium ferrocyanide powder dispersed in the epoxy resin. Vertical scale is
the radiation transmission normalized to unity far from any resonance. Horizontal
scale is the velocity of the source in mm/sec. Piezo transducer was fed by RF voltage
with 12.27 MHz frequency. The voltage is 0 V (a), 2 V (b), 4V (b), and 12 V (c).
Experimental spectra are shown by blue dots and theoretical fitting by Eq. (\ref{Eq4})
is depicted by red solid line.}%
\label{fig:1}%
\end{figure}
\begin{figure}[ptb]
\resizebox{0.5\textwidth}{!}{\includegraphics{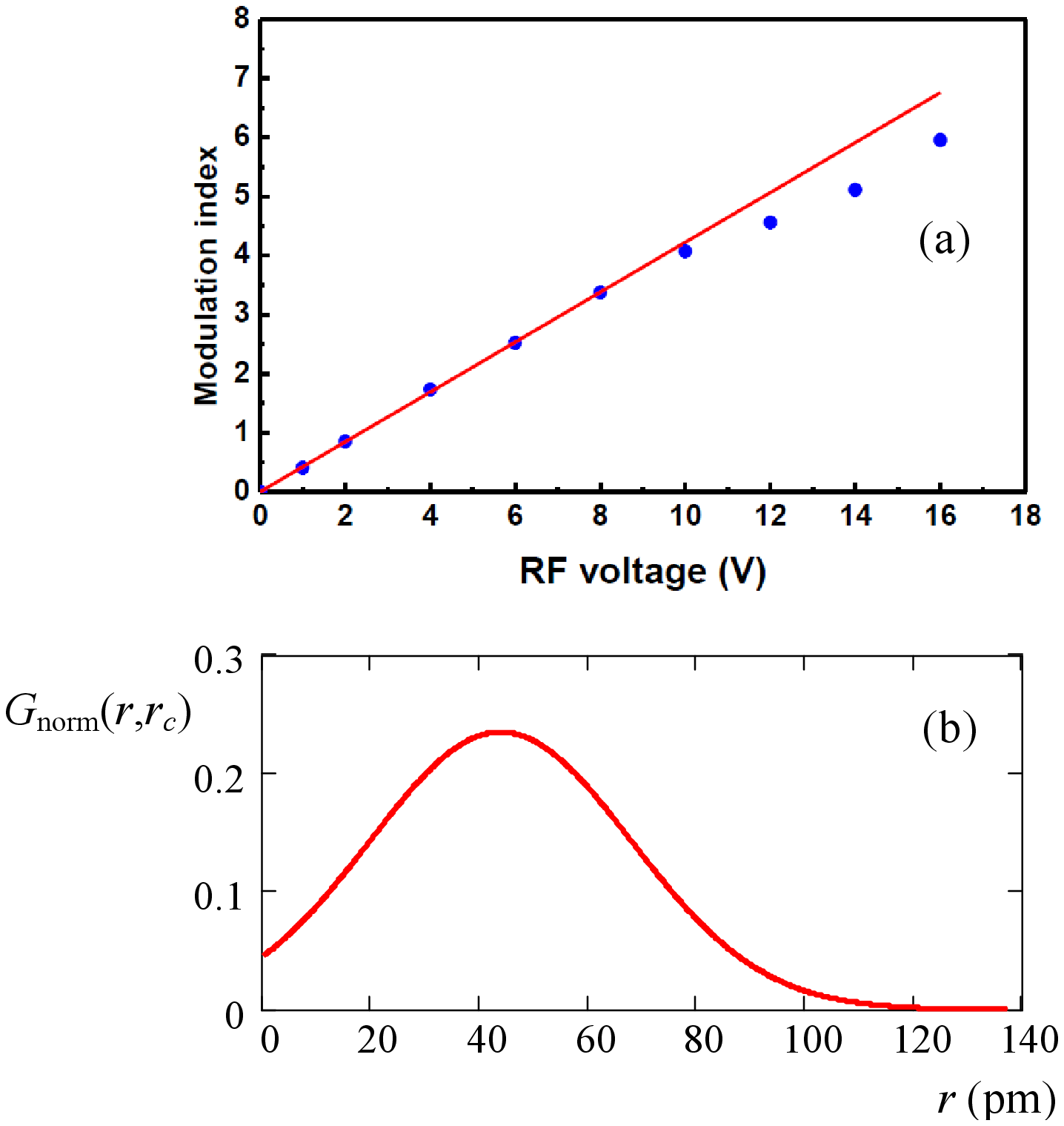}}\caption{(a) Dependence of
the mean value of the modulation index on the voltage of the RF generator.
(b) Distribution of the vibration amplitudes of iron nuclei in epoxy resin,
$G_{\textrm{norm}}(r,r_{c})$, for 8 V. Horizontal scale is in picometers.}%
\label{fig:2}%
\end{figure}

We also made experiments using the led mask with a small round hole 0.6 mm
in diameter. In the previously reported experiment \cite{SV} this mask allowed
to observe spectra of different small parts of the vibrated SS foil. It was found
that within an area 0.6 mm in diameter the foil vibrates almost uniformly.
Meanwhile, the vibration amplitudes at the foil center and at its edges strongly
differ. Periodical displacements at the edges almost 1.5 times larger that at the
center. This can be explained by the small density of the polymer piezo transducer
(PVDF), creating periodical displacements, with respect to the density of SS
foil of approximately the same thickness as PVDF film. Therefore, the film is loaded
larger at its center and smaller at the edges. This explains the difference of
the foil spectra at the foil center and at its edges. Moreover, SS foil was glued
to PVDF film by the polymerized epoxy glue. Since SS foil is rigid against extension,
extra resistance appears to the PVDF film deformation in the longitudinal direction,
i.e., to the stretching and shortening of the polymer chains, aligned along
the film and producing the film surface vibration due to lengthening/shortening.

The absorber, which we use in this paper, is a spot of epoxy resin (5$\times$5 mm)
containing colloidal particles of sodium ferrocyanide. The density of this absorber
is comparable with the density of the polymer piezo transducer. Moreover,
epoxy resin without hardener should not create a resistance to the PVDF film
movements. Therefore, we expected that the center and the edges of the absorber
will vibrate with the same amplitude. Scanning of the led mask with the small hole
in two orthogonal directions showed that from point to point spectra do not
change within the accuracy of the experiment (see Fig. 3 showing the dependence
of the modulation index on position of the hole along the absorber). Meanwhile,
in each small area, covered by the hole, nuclei vibrate with amplitudes distributed
according the function shown in Fig 2(b). This is possible if the ultrasound,
induced by PVDF film, decays along the direction of gamma-radiation propagation.
The particles, which are located close to the pizo transducer, i.e., at the bottom
of the epoxy resin, vibrate with lager amplitude. With distance of the layer
of particles from the contact of resin with PVDF the vibration amplitude decreases.
This kind of distribution of the vibration amplitudes reflects a decay of
ultrasound in the epoxy resin.

In conclusion we summarize our results. It is found that a single line of $^{57}$Fe
in colloidal particles is transformed into a central line with many satellites if the
particles are vibrated by ultrasound piezo transducer. The spacing between the
nearest satellites is equal to the vibration frequency. The number of the satellites
depends on the vibration amplitude of nuclei and its distribution in the
area of the absorber irradiated by gamma-radiation. The distribution function of
the vibration amplitudes derived from comparison of experimental spectra with the
theoretical predictions indicates that mechanical vibration of particles decay
in the viscous medium. The obtained results demonstrate the possibility of application
of gamma-resonance spectroscopy for studying the vibrational dynamics of
colloidal particles immersed into a viscous medium.
\begin{figure}[ptb]
\resizebox{0.5\textwidth}{!}{\includegraphics{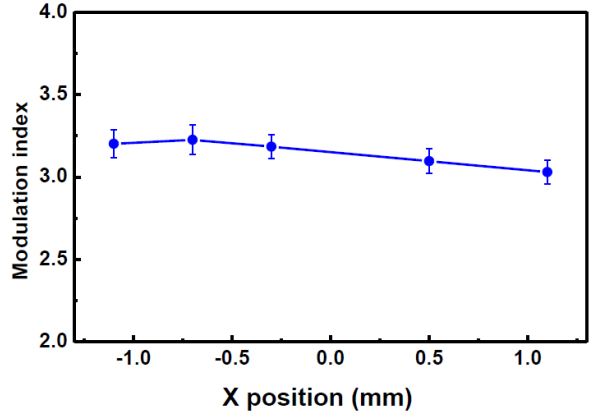}}\caption{Dependence of
the modulation index for the same RF voltage (8 V) on the position of the hole
in the mask with respect to the absorber center (in millimeters).}%
\label{fig:3}%
\end{figure}

\section{Acknowledgements}

The experimental work is supported by Russian Foundation for Basic Research
(Grant No. 18-02-00845-a) and partially by the Program of Competitive Growth
of Kazan Federal University funded by the Russian Government.

\end{document}